# CALIBRATION OF THE TRANSPORT PARAMETERS OF A LOCAL PROBLEM OF WATER QUALITY IN IGAPÓ I LAKE

**Neyva M.L. Romeiro*†, Rigoberto G.S.Castro††, Eliandro R. Cirilo† and Paulo L. Natti†**

† Universidade Estadual de Londrina (DMAT/UEL)
Rodovia Celso Garcia Cid, Pr 445 Km 380, C.P. 6001, CEP 86051-990, Londrina, Paraná, Brasil
e-mail: {nromeiro,ercirilo,plnatti}@uel.br, web page: http://www.mat.uel.br

†† Universidade Estadual do Norte Fluminense (LCMAT/UENF)
Av. Alberto Lamego, 200, CEP 28015-620, Campos dos Goytacazes, Rio de janeiro, Brasil
e-mail: sanabria@uenf.br, web page: http://www.uenf.br

**Summary.** *The calibration of a model refers to the process by which one can estimate some parameters by comparisons with observed data. Due to the dynamical nature of the environment, variations between predicted and observed values occur. Thus, the environmental parameters may vary due to random temperature changes, time of discharge flow, time of the day, and other conditions. Such variations can be minimized by identifying and optimizing some parameters of the transport model, like the values of diffusion coefficients in x and y directions and the kinetic parameter that describes the process of removing pollutants. This paper presents results concerning the calibration of transport parameters for two-dimensional problems of water quality (fecal coliform control) at Igapó I Lake, located in Londrina, Paraná, Brazil. Thus, the convection-diffusion-reaction equation, which describes mathematically the process studied in this work, is resolved by a semi-discrete finite element method (SUPG) which combines finite differences in time and finite elements in space.*

## 1 INTRODUCTION

The concern with water pollution has motivated the development of practices for the calibration of mathematical models by means of numeric simulation of flow and transport. Calibration refers to the process by which parameters estimates are made through observed data. Calibrated and validated models, in relation to the current conditions of a water body, may be used as water quality monitoring tools.

In this work, the distribution of pollution concentration is simulated at Igapó I Lake located in Londrina, Paraná, Brazil, by using fecal coliform concentration as a representative substance of water quality. Fecal coliforms bacteria are found in the intestinal tract of humans and other hot-blooded animals. Although fecal coliform bacteria are not necessarily dangerous to men, their high concentrations in water bodies indicate hazards for the human health[1]. Thus, the monitoring of such concentrations consists of a method to identify contamination magnitude, and a very well calibrated mathematical model may be used to describe such pollution.

For the calibration of the transport model proposed in this study, three values for the longitudinal diffusion coefficient, direction $x$, and five values for the transversal diffusion coefficient, direction $y$, are used, considering as fixed the kinetic parameter that describes the pollutant removal process.

The convection-diffusion-reaction equation, which models the fecal coliform concentrations at Igapó I Lake, is resolved by a semi-discrete finite elements method (SUPG), which combines finite differences in time and finite elements in space[2].

## 2 NUMERICAL MODEL

### 2.1 Governing equation

The 2D model for the convection-diffusion-reaction equation for a pollutant, with spatial variable $\mathbf{x} = (x, y)$, can be written as

$$\frac{\partial C}{\partial t} + \mathbf{q} \cdot \nabla C - \nabla \cdot (\mathbf{D} \cdot \nabla C) + KC = 0, \quad \mathbf{x} \in \Omega, \quad t \in [0, T] \tag{1}$$

where $C$ is the concentration; $t$ is the temporal variable; $\mathbf{q} = (u, v)$ describes fluid velocity vector in $x$ and $y$ directions, respectively; $\mathbf{D} = \begin{bmatrix} D_x & 0 \\ 0 & D_y \end{bmatrix}$ is the diffusion matrix, where $D_x$ is the longitudinal diffusion coefficient and $D_y$ is the transversal diffusion coefficient, $K$ is the linear kinetic parameter that describes the pollutant removal process; $\Omega$ is a bounded problem domain in the 2D space, and $T$ is a time defined in the problem. The initial condition of this model will be considered zero and in the boundary described by

$$C_B = C_B(t), \quad t > 0. \tag{2}$$

To resolve the model (1)-(2), a semi-discrete finite elements method (SUPG) is used, which combines finite differences in time and finite elements in space.

### 2.2 Finite element formulation

The methodology named semi-discrete finite elements is characterized by the combination of distinct approximations for the spatial and temporal variables of equation (1), where the temporal derivative is approximated by an implicit scheme of finite differences and the spatial solution is obtained by using the stabilized finite elements method, SUPG (Streamline Upwind Petrov-Galerkin). This methodology was used successfully in the search for solutions for transport problems with convective character[3,4], therefore justifying its use in this research as well.

Given $\Omega$ a spatial domain with boundary $\Gamma$. Subdividing the domain in *nel* elements $\Omega^e$, where $\Omega = \bigcup_{e=1}^{nel} \Omega^e$, $\Omega^i \cap \Omega^j = \emptyset$ para $i \neq j$. The subspaces of finite elements, at time $t = t_n$, are given by

$$\hat{\mathbf{U}}_n^h \equiv \left\{ w^h; \ w^h \in \left(C^0(\Omega)\right)^3; \ w^h\big|_{\Omega^e} \in \left(P^k(\Omega^e)\right)^3; \ w^h\big|_\Gamma = \mathbf{0} \right\} \tag{3}$$

$$\mathbf{U}_n^h \equiv \left\{ C^h(.,t); \ C^h \in \left(C^0(\Omega)\right)^3; \ C^h\big|_{\Omega^e} \in \left(P^k(\Omega^e)\right)^3; \ C^h\big|_\Gamma = C_B \right\}, \tag{4}$$

where $\hat{\mathbf{U}}_n^h$ is the subspace of the ponderation functions, $\mathbf{U}_n^h$ is the subspace of the admissible functions, $P^k$ is the set of polynomial functions of degree lower than or equal $k$, $C^0(\Omega)$ is the set of continuous functions and $C_B$ are the prescribed boundary conditions.

The stabilized semi-discrete formulation of the finite elements method consists of: given $C(t_0)$, for each time $t_n$, $n = 1, 2, ...$, find $C^h \in \mathbf{U}_n^h$, so that $\forall \, w^h \in \hat{\mathbf{U}}_n^h$, the following variational problem is satisfied

$$\int_\Omega R^h \cdot w^h d\Omega + \sum_{e=1}^{nel} \left[ \int_\Omega R^h \cdot (\tau \mathbf{q} \cdot \nabla w^h) d\Omega \right] = 0, \tag{5}$$

where $R^h = \dfrac{\partial C^h}{\partial t} + \mathbf{q} \cdot \nabla C^h - \nabla \cdot (\mathbf{D} \cdot \nabla C^h) + K C^h$ is the residue associated to the approximate solution $C^h$. In equation (5), the first integral corresponds to Galerkin's formulation and the second integral is the term SUPG (Streamline Upwind Petrov-Galerkin). The stabilization term is given by $\tau = \dfrac{1}{2} \dfrac{\alpha \, h^e}{\|\mathbf{q}\|}$, where $\alpha = \min\left(\dfrac{Pe}{3}, 1\right)$ is an upwind term, $Pe$ is the Peclet number of element $\Omega^e$, calculated as $Pe = \dfrac{1}{2} \dfrac{\|\mathbf{q}\| h^e}{|\mathbf{q}^T \mathbf{D} \mathbf{q}|}$, with $h^e = \sqrt{A}$ as the characteristic size of element $\Omega^e$ and $A$ the area of element. Considering $C^h(\mathbf{x}, t) \approx \sum_{j=1}^{nel} C_j(t) \, w_j^h(\mathbf{x})$ in equation (5), the following system of linear ordinary differential equation is obtained $\mathbf{M} \dot{\mathbf{C}} + \mathbf{K} \mathbf{C} = \mathbf{F}$, where $\mathbf{C} = (C_1, C_2, ..., C_{nel})^T$, $\mathbf{M}$ is the mass matrix, $\mathbf{K}$ is the rigidity matrix and $\mathbf{F}$ is the forces vector. The previous system is resolved by using a scheme of finite differences.

## 3 ANALYSIS OF POLLUTANT TRANSPORT AT IGAPÓ I LAKE

### 3.1 Location of Igapó I Lake

Igapó Lake, located in Londrina, Paraná, Brazil, is situated in the microbasin of Cambé Stream, whose spring is in the town of Cambé, approximately 10 km from the city of Londrina, in the State of Paraná. After the spring, it flows to the west crossing all the southern area of Londrina, gathering many streams along its way. The Lake is subdivided into: Igapó I, II, III and IV, as shown in Figure 1. They were designed in 1957 as a solution for the Cambé Stream drainage problem.

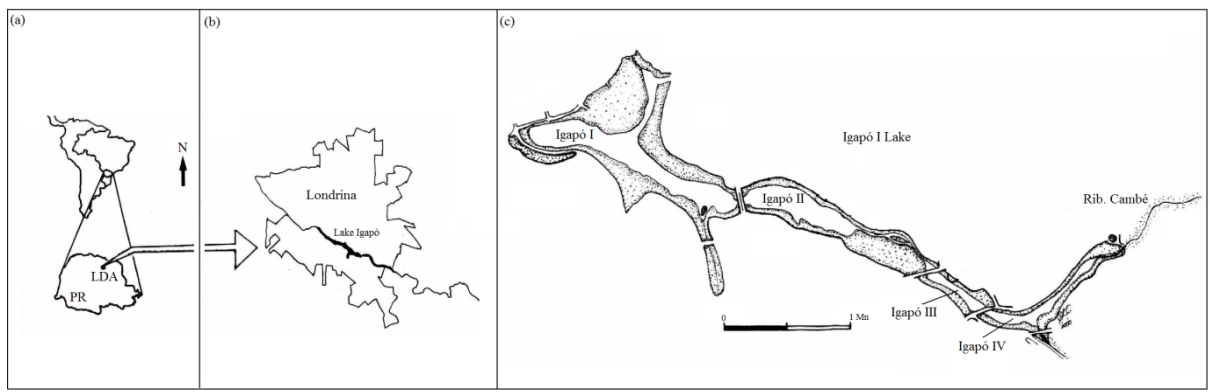

Figure 1: (a) South America, Brazil and Paraná State, (b) City of Londrina with Igapó Lake highlighted, (c) Map of Igapó Lake, Londrina, Paraná.

Because it is located near the central area of the city of Londrina, Igapó I Lake receives the discharge of non-treated pollutants in its waters, besides the discharge of pollutants from Lakes IV, III e II, which pollute the Lake I. Thus, the present work analyses, by means of a numeric simulation, the impact of pollutant discharge in Igapó I Lake. The longitudinal length of the Lake is approximately 1.8 Km. As observed in Figure 2a, the water flows from Igapó II Lake into Igapó I Lake when crossing Higienópolis Avenue, which characterizes its entrance. In the left bank there is undergrowth, as well as an input channel. The right bank is split in private properties and contains another input channel. The exit is a physical dam and the water flow is controlled by water pipes and ramps.

The computational grid for the numerical simulation, consisting of 23.124 triangular elements and 11.886 nodes, was build after an aerophotogrametric study by using a non-geocentric reference system SAD 69 (South American Datum 1969), which produced the coordinates ($x$, $y$) of the left and right banks at Igapó I. Figure 2b presents a computational network and location of the data for the numerical simulation. In point (c), the arrows illustrate the water entrance of the lake, while in (d) has the numerical model verification point (arrow near to the dam). The banks were obtained by the parametric polynomial interpolation cubic spline method[5,6]. Observe that in the computational grid in Figure 2b, the input channels have been deleted.

By mathematically modelling the flow with incompressible fluid by means of the Navier-Stokes equation system and pressure equation, submitted to initial and boundary conditions, the velocity field $\mathbf{q}=(u,v)$ in the computational grid was numerically obtained[6]. By inserting this velocity field as input data in the transport model (1), the concentrations of fecal coliforms are numerically obtained via semi-discrete finite elements formulation

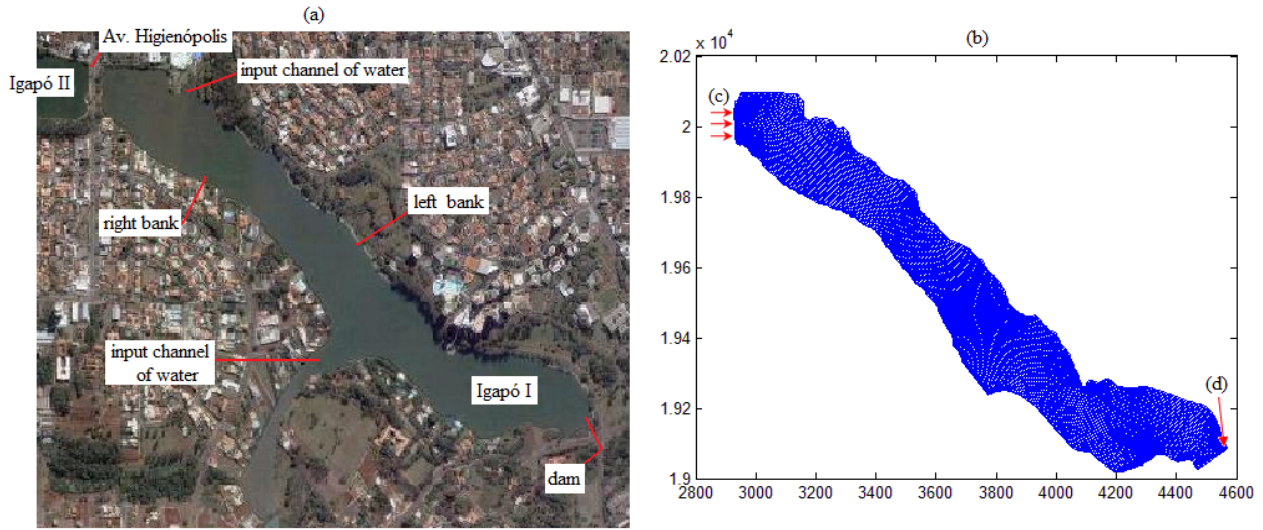

Figure 2: (a) Physical domain of Igapó I Lake, and (b) Computational network, where (c) Lake entry, considered as boundary condition in the model, (d) Verification point of numerical simulation.

The data collected[7] on August 19, 2008 provided in Lake entrance, point (c) in Figures (2b), the values 11000 MPN/100ml as Most Probable Number (MPN) of fecal coliforms per one hundred milliliters of sampled water (MPN/100 ml), and 2100 MPN/100ml for its exit, point (d) in Figure (2b).

For the linear decay coefficient $K$, the literature[8] indicates daily average values of $K$ in the range of 0.5 to $2\,\mathrm{d}^{-1}$. The calibrated value of $K$ for the August 2008 was used as $K = 0.02083\,\mathrm{h}^{-1}$. As for the calibration of the longitudinal $D_x$, and transversal $D_y$ diffusion coefficients, values $D_x = 0.1$, $1.0$, $10.0$ m$^2$/h and $D_y = D_x \alpha_i$, m$^2$/h, where $\alpha_i = 10^{i-3}$, $i = 0,\ldots,4$, are taken, as described in Table 1.

| Values of $\alpha$ | $D_y = 0.1\alpha_i$ | $D_y = 1.0\alpha_i$ | $D_y = 10.0\alpha_i$ |
|---|---|---|---|
| $\alpha_0 = 0.001$ | 0.0001 | 0.001 | 0.01 |
| $\alpha_1 = 0.01$ | 0.001 | 0.01 | 0.1 |
| $\alpha_2 = 0.1$ | 0.01 | 0.1 | 1.0 |
| $\alpha_3 = 1.0$ | 0.1 | 1.0 | 10.0 |
| $\alpha_4 = 10.0$ | 1.0 | 10.0 | 100.0 |

Table 1: Possible longitudinal and transversal diffusion coefficients.

The results of the numerical simulations for the concentrations of fecal coliforms, during 500 time steps with $\Delta t = 0.4$ h, when the steady concentration was obtained, checked in the verification point, can be observed in Table 2.

| Values of $\alpha$ | Concentrations for | | |
|---|---|---|---|
| | $D_y = 0.1\alpha_i$ | $D_y = 1.0\alpha_i$ | $D_y = 10.0\alpha_i$ |
| $\alpha_0 = 0.001$ | 2140.1 | 2120.8 | 2465.2 |
| $\alpha_1 = 0.01$ | 2129.6 | 2186.9 | 2466.6 |
| $\alpha_2 = 0.1$ | 2166.3 | 2214.4 | 2467.6 |
| $\alpha_3 = 1.0$ | 2166.7 | 2215.9 | 2545.5 |
| $\alpha_4 = 10.0$ | 2197.8 | 2193.1 | 3376.4 |

Table 2: Numerical simulations of fecal coliform concentration in the verification point at Igapó I Lake in function of longitudinal and transversal diffusion coefficients.

The values of $D_x$ and $D_y$ are chosen so as to ensure the best adjustment of fecal coliform concentration in the verification point (near to the dam, the left bank). In this work, the transversal diffusion coefficient value that was closer to the fecal coliform concentration value observed was at $\alpha_1$ for $D_x = 0.1$ m²/h and at $\alpha_0$ for $D_x = 1.0$ and $D_x = 10.0$ m²/h. Such results are presented in Table 2. The colored cells indicate the best calibrated values to the diffusion coefficients. It is also observed that the variation of the results for $D_x = 0.1$ and $D_x = 1.0$ m²/h do not differ significantly, therefore, the Figure 3 presents the results of the simulations at Igapó I Lake for the situations $D_x = 1.0$ m²/h with $D_y = 0.1$ m²/h, and $D_x = 10.0$ m²/h with $D_y = 0.1$ m²/h.

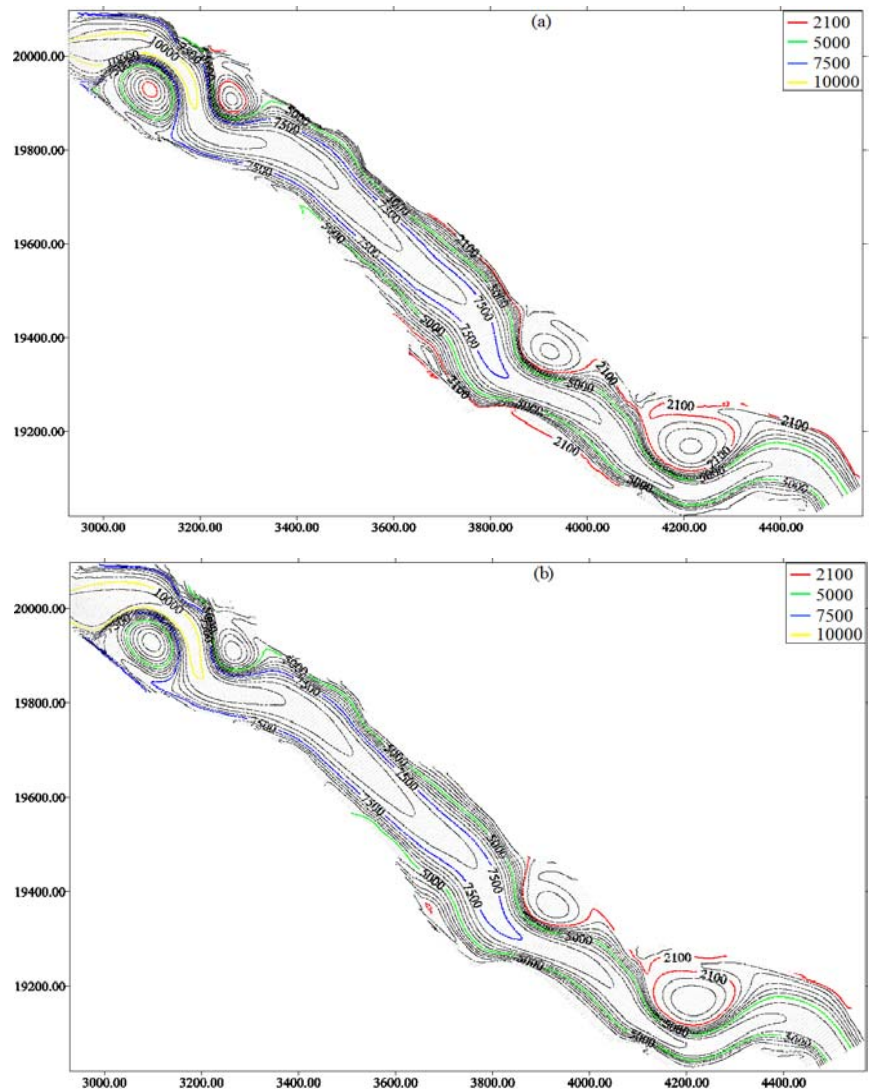

Figure 3: Results of simulations at Igapó I Lake, when: (a) $D_x = 1.0$ m$^2$/h with $D_y = 0.001$ m$^2$/h, and (b) $D_x = 10.0$ m$^2$/h with $D_y = 0.01$ m$^2$/h. The variation in fecal coliform concentrations is described by the contour lines: (—) 2100 MPN/100ml, (—) 5000 MPN/100ml, (—) 7500 MPN/100ml, and (—) 10000 MPN/ml, respectively.

## 4 CONCLUSIONS

A bidimentional advection-diffusion-reaction model was used to simulate the transport of pollutants at Igapó I Lake, located in Londrina, Paraná, Brazil. Fecal coliforms were chosen as a representative substance to assess water quality in the lake. Fecal coliform concentration measurements in samples collected on August 19, 2008 were considered in the simulations. Finally, diffusion parameters (longitudinal and transversal diffusion coefficients) were calibrated. The results presented in this study show that this numerical model can adequately explain the distribution of fecal coliforms at Igapó I Lake, as described by the contour lines in Figures 3a and 3b.

## ACKNOWLEDGEMENTS

The first author acknowledges CNPq-Brazil (National Council for Scientific and Technological Development) for the financial support to this research.